\documentclass[numreferences]{kluwer}
\newcommand{\bra}[1]{\langle#1|}
\newcommand{\ket}[1]{|#1\rangle}
\begin{document}
\begin{article}
\begin{opening}
\title{The Frobenius formalism in Galois quantum systems}

\author{A. \surname{Vourdas}}
 
\institute{Department of Computing,
University of Bradford, 
Bradford BD7 1DP, United Kingdom}

\runningtitle{Galois quantum systems}
\runningauthor{Vourdas}

\begin{abstract}
Quantum systems in which the position and momentum take values in the ring ${\cal Z}_d$ and which are described
with $d$-dimensional Hilbert space, are considered.
When $d$ is the power of a prime,
the position and momentum take values  in the Galois field $GF(p^ \ell)$, the position-momentum phase space
is a finite geometry and the corresponding `Galois quantum systems' have stronger properties.
The study of these systems uses ideas from the subject
of field extension in the context of quantum mechanics.
The  Frobenius automorphism in Galois fields leads to Frobenius subspaces and
Frobenius transformations in Galois quantum systems.
Links between the Frobenius formalism and Riemann surfaces, are discussed. 
\end{abstract}


\classification{JEL codes}{81S30; 11S20}
\end{opening}

\section{Introduction}

Phase space methods play an important role in quantum mechanics.
In the case of a harmonic oscillator, both the position $x$ and the momentum $p$ take values in $R$
(real numbers) and the position-momentum phase space is the plane $R\times R$.
The state $\ket{f}$ of the system is described with the wavefunction $f(x)$ in the x-representation,
or with the wavefunction ${\tilde f}(p)$ in the p-representation. These two wavefunctions are related through the Fourier transform:
\begin{eqnarray}
&&{\tilde f}(p)=(2\pi )^{-1/2}\int dx f(x)\exp (-ixp)\nonumber\\
&&\int dx |f(x)|^2=\int dp |{\tilde f}(p)|^2=1
\end{eqnarray}
The states $\ket{f}$ belong in an infinite-dimensional Hilbert space.
There are two important classes of transformations in the phase space $R\times R$.
The first is displacements and they are associated with the Heisenberg-Weyl group;
and the second is symplectic transformations and they are associated with the
symplectic group $Sp(2,R)$.

In this article we are interested in quantum systems described with finite-dimensional Hilbert spaces.
These systems have been studied originally by Weyl\cite{1} and Schwinger\cite{2}, and later by many authors \cite{3,4,5,6,7,8,9,10}.
A review of the subject with an extensive list of references has been given in \cite{rev}. 
In this case the position and momentum take values in the ring ${\cal Z}_d$ (the
integers modulo $d$) and the phase space of the system is the toroidal lattice ${\cal Z}_d\times {\cal Z}_d$.
Displacements in this phase space are discrete and form a Heisenberg-Weyl group. 

The next step is to try to define symplectic transformations.
We note however that the ${\cal Z}_d\times {\cal Z}_d$ phase space is in general a collection of points 
with no geometrical structure and we cannot define symplectic transformations.
Consequently the properties of such systems are weaker in comparison to the harmonic oscillator.
The root of these difficulties is that ${\cal Z}_d$ is a ring.
However when the dimension $d$ of the Hilbert space of the system is the power of a prime number $p$ (i.e., $d=p^ \ell$) 
the ${\cal Z}_d$ (with appropriate multiplication rule) becomes the Galois field $GF(p^\ell)$.
In this case the phase space is a finite geometry \cite{fg} and translations and rotations are well defined and they form groups.
In this case we can define the group of symplectic transformations $Sp(2,GF(p^\ell))$\cite{8}.

Another problem which leads to similar ideas through another route, is to find mutually unbiased bases (orthonormal bases $\ket{a_i}$ and 
$\ket{b_j}$ such that $|\bra {a_i}b_j\rangle |^2=d^{-1}$)\cite{m0,m1,m2,m3,m4,m5,m6,m7,m8,m9,m10}. 
It is known that the number of such bases cannot exceed $d+1$; and it is also 
known that for systems where $d$ is the power of a prime, the number of such bases is indeed $d+1$.
Related to this is also the so-called `mean king's problem'\cite{mkp1,mkp2}.

Galois fields play a central role in classical coding and
these techniques could be transfered in quantum information processing.
Work with Galois fields in the context of quantum coding has been reported in \cite{c1,c2,c3}.

In summary, Galois fields in finite quantum systems have been introduced either 
in order to have well defined symplectic transformations;
or in the context of mutually unbiased bases; or in the context of quantum coding.  
In this paper we are interested  
in the use of Galois fields in quantum mechanics, as a subject in its own right rather than as an application driven problem.
Most of the material presented here has appeared in the literature from a different point of view. 
The aim of this review article is to present these ideas with a different emphasis and
promote the interaction between Galois fields and quantum mechanics.

We transfer the concept of field extension in the context of Hilbert spaces.
We start with a $p$-dimensional Hilbert space ${\cal H}$ (where $p$ is an odd prime number)
and through tensor product of $\ell$ such spaces, we construct
a $p^\ell$-dimensional Hilbert space $H$.
In this space the Fourier transform $F$ is defined in terms of additive characters in $GF(p^\ell )$ and 
is different from ${\cal F}\otimes...\otimes {\cal F}$, where ${\cal F}$ is the Fourier transform in ${\cal H}$.
The Fourier transform $F$ is very important because it provides the `Galois identity' to these systems.
We then develop concepts like Galois conjugate position (and momentum) states; Frobenius subspaces;
Frobenius transformations, etc.
Frobenius automorphisms play an important role in Galois fields; and here we develop analogous concepts in Hilbert spaces
in the context of quantum mechanics (and in the related area of applied harmonic analysis).
We also explain that there are interesting connections between the Frobenius formalism in Galois theory and Riemann surfaces. 

In section 2 we present some ideas from the theory of Galois fields, in the language of matrices.
In section 3 we discuss the basic theory of general systems with finite Hilbert space.
Galois quantum systems are discussed in section 4.
An important part of Galois theory is the Frobenius transformations and their
implications in the present context are studied in section 5.
Links between the Frobenius formalism in a quantum mechanical context and Riemann surfaces are discussed in section 6.
We conclude in section 7 with a discussion of our results.

.
\section{Galois fields}

We consider the Galois field $GF(p^\ell)$.
Its elements can be written as polynomials:
\begin{eqnarray}\label{098}
\alpha =\alpha _0+ \alpha _1\epsilon +...+\alpha _{\ell-1}\epsilon ^{\ell-1}
;\;\;\;\;\;\;\alpha _0, \alpha _1,...,\alpha _{\ell-1}\in {\cal Z}_p 
\end{eqnarray}
These polynomials are defined modulo an irreducible polynomial of degree $\ell$:
\begin{eqnarray}\label{cha}
P(\epsilon)\equiv c _0+ c _1\epsilon +...+c _{\ell-1}\epsilon ^{\ell-1}+\epsilon ^\ell
;\;\;\;\;\;\;c _0, c _1,...,c _{\ell-1}\in {\cal Z}_p 
\end{eqnarray}
Different irreducible polynomials of the same degree $\ell$ lead to isomorphic finite fields. 
Results of practical calculations do depend on the choice of the 
irreducible polynomial, but different choices lead to isomorphic results.

The powers $\alpha$, $\alpha ^p$,...,$\alpha ^{p^{\ell-1}}$ are Galois conjugates.
The elements of the base field ${\cal Z}_p$ are Galois self-conjugates.
The trace of $\alpha$ is the sum of all its conjugates:
\begin{eqnarray}\label{777}
{\rm Tr}(\alpha)=\alpha +\alpha ^p+...+\alpha ^{p^{\ell-1}};\;\;\;\;\;\;\;{\rm Tr}(\alpha)\in {\cal Z}_p
\end{eqnarray}
All conjugates have the same trace.

We consider a nonzero constant $\mathfrak h$ in $GF(p^\ell)$ which we call `inverse Planck's constant'
because later will play such a role in quantum systems. 
For practical calculations we introduce the:
\begin{eqnarray}
{\cal E}_ \lambda & \equiv &{\rm Tr} ({\mathfrak h}\epsilon ^\lambda);\;\;\;\;\;\;\;{\cal E}_\lambda \in{\cal Z}_p
\end{eqnarray}
We also introduce the following
$\ell\times \ell$ matrices with elements in ${\cal Z}_p$:
\begin{eqnarray}
g_{\lambda \kappa}\equiv{\cal E}_{\lambda +\kappa};\;\;\;\;\;G_{\kappa \lambda}\equiv (g^{-1})_{\kappa \lambda};\;\;\;\;\;G_{\kappa \lambda}\in {\cal Z}_p
\end{eqnarray}
We note that the determinant of $g$ is non-zero and therefore its inverse $G$ exists.

The $GF(p^ \ell)$ can be regarded as a $\ell$-dimensional vector space with the $1,\epsilon, \epsilon ^2,...,\epsilon ^{\ell -1}$
as a basis. We can change this basis into a different basis, and for later use we introduce the dual basis 
$E_0, E_1,...,E_{\ell -1}$, as follows:
\begin{eqnarray}
E_{\kappa}=\sum _{\lambda } G_{\kappa \lambda}\epsilon ^{\lambda };\;\;\;\;\;\;\;\;
{\rm Tr}({\mathfrak h}\epsilon ^\kappa E_{\lambda})=\delta _{\kappa \lambda}
\end{eqnarray}
A number $\alpha \in GF(p^\ell)$ can be expressed in the two bases as:
\begin{eqnarray}\label{kkk}
\alpha &=&\sum _{\lambda=0}^{\ell-1}\alpha _ {\lambda} \epsilon ^{\lambda}=
\sum _{\lambda=0}^{\ell-1}{\bar \alpha} _\lambda E_{\lambda }\nonumber\\
\alpha _\lambda &=&{\rm Tr}[{\mathfrak h}\alpha E_{\lambda }];\;\;\;\;\;
{\bar \alpha} _{\lambda} ={\rm Tr}[{\mathfrak h}\alpha {\epsilon} ^\lambda ]
\end{eqnarray}
We refer to $\alpha _{\lambda }$ and ${\bar \alpha} _{\lambda }$ as the components and dual components of $\alpha$, correspondingly.
They are related as follows:
\begin{eqnarray} 
\alpha _{\lambda }=\sum _{\kappa} G_{ \lambda \kappa}{\bar \alpha} _{\kappa};\;\;\;\;\;\;
{\bar \alpha} _{\lambda }=\sum _{\kappa} g_{ \lambda \kappa}\alpha _{\kappa}
\end{eqnarray}

The trace of ${\mathfrak h}\alpha$ is given by
\begin{eqnarray}
{\rm Tr}({\mathfrak h}\alpha )&=&\sum _{\lambda=0}^{\ell -1}\alpha _\lambda {\cal E}_{\lambda}
\end{eqnarray}
 If $\beta$ is another number in $GF(p^\ell)$
\begin{eqnarray}
\beta&=&\sum _{\lambda=0}^{\ell-1}\beta _ {\lambda} \epsilon ^{\lambda}=
\sum _{\lambda=0}^{\ell-1}{\bar \beta} _\lambda E_{\lambda }
\end{eqnarray}
the trace of  ${\mathfrak h}\alpha \beta $ is given by
\begin{eqnarray}\label{102}
{\rm Tr}({\mathfrak h}\alpha \beta)&=&\sum _{\lambda,\kappa}g_{\lambda \kappa}\alpha _\lambda \beta _\kappa =
\sum _{\lambda,\kappa}G_{\lambda \kappa}{\bar \alpha} _\lambda {\bar \beta} _\kappa\nonumber\\&=&
\sum _{\lambda} \alpha _\lambda {\bar \beta}_\lambda=\sum _{\lambda} {\bar \alpha}_\lambda \beta _\lambda
\end{eqnarray}

\subsection{Galois conjugates}

For practical calculations of the Galois conjugates we introduce the $\ell \times \ell$ matrix ${\cal C}$
with elements in ${\cal Z}_p$, through the relations:
\begin{eqnarray}
\epsilon ^{\mu p}=\sum _{\kappa =0}^{\ell -1}\epsilon ^\kappa {\cal C}_{\kappa \mu}
\end{eqnarray}
Here $\kappa , \mu $ take values from $0$ to $\ell -1$.
We can show that more generally
\begin{eqnarray}
\epsilon ^{\mu p ^\lambda }=\sum _{\kappa =0}^{\ell -1}\epsilon ^\kappa ({\cal C}^\lambda )_{\kappa \mu}
\end{eqnarray}
where $\lambda $ take values from $0$ to $\ell -1$.
For $\lambda =0$ we have ${\cal C} ^0 ={\bf 1}$. Also
\begin{eqnarray}
{\cal C}^\ell ={\bf 1};\;\;\;\;\;\;{\cal C}_{\kappa 0}=\delta (\kappa ,0)
\end{eqnarray}
where $\delta $ is the Kronecker delta.
We can now express the conjugates of the arbitrary number $\alpha$ of Eq.(\ref{098}) as
\begin{eqnarray}\label{777}
\alpha ^{p^ \lambda}=\sum _{\kappa ,\mu}\epsilon ^\kappa ({\cal C} ^\lambda )_{\kappa \mu}\alpha _{\mu}
\end{eqnarray}

The Frobenius map 
\begin{eqnarray}\label{Fro}
\sigma :\alpha \;\rightarrow \; \alpha ^p
\end{eqnarray}
defines an automorphism in $GF(p^\ell)$.
It maps the Galois conjugates to each other and
leaves all elements of the base field ${\cal Z}_p$ fixed.
The Frobenius map can be written in terms of the components of $\alpha$ and $\alpha ^p$ as:
\begin{eqnarray}
\sigma : \alpha _{\kappa}\;\rightarrow \;\sum _{\mu} {\cal C} _{\kappa \mu}\alpha _{\mu} 
\end{eqnarray}

\subsection{Characters}
Below we will use the complex-valued function 
\begin{equation}\label{ppp}
\chi (\alpha)=\omega [{\rm Tr}(\alpha)];\;\;\;\;\;\;\;\;\omega= \exp \left (i\frac{2\pi}{p}\right )
\end{equation}
This is an additive character in $GF(p^\ell)$:
\begin{equation}\label{pp}
\chi (\alpha)\chi (\beta)=\chi (\alpha +\beta);\;\;\;\;\;\;\;\;\alpha ,\beta \in GF(p^\ell)
\end{equation}
We can easily show that for $n,m,r \in GF(p^\ell)$:
\begin{eqnarray}\label{23}	
\frac{1}{p^\ell}\sum _n \omega \left [{\rm Tr}(nm-nr)\right ]=\delta (m,r)
\end{eqnarray}
A more general relation is
\begin{eqnarray}\label{24}	
\frac{1}{p^\ell}\sum _n \omega \left [{\rm Tr}(nm-n^{p^\lambda}r)\right ]=\delta \left (m,r^{p^{\ell-\lambda}}\right)=
\delta \left (m^{p^{\lambda}},r\right )
\end{eqnarray}
We note that in terms of the components of the $m^{p^{\lambda}},r$
\begin{eqnarray}
\delta \left (m^{p^{\lambda}},r\right )=\prod _{\kappa =0}^{\ell -1}\delta \left (\sum _{\mu}({\cal C} ^\lambda )_{\kappa \mu}m _{\mu},r_\kappa \right )
\end{eqnarray}

\subsection{Example}\label{AAA1}

We consider the Galois fields $GF(9)$ and calculate the quantities defined above
for the irreducible polynomial $\epsilon ^2+\epsilon +2$. We find that
\begin{eqnarray}
{\cal C}=
\left (
\begin{array}{cc}
1&-1\\
0&-1
\end{array}
\right)
\end{eqnarray}
We also choose ${\mathfrak h}=1$ and find that
\begin{equation}
{\cal E}_0=-1;\;\;\;\;\;{\cal E}_1=-1;\;\;\;\;\;{\cal E}_2=0
\end{equation}
In this case
\begin{eqnarray}
g=
\left (
\begin{array}{cc}
-1&-1\\
-1&0
\end{array}
\right);\;\;\;\;\;
G=
\left (
\begin{array}{cc}
0&-1\\
-1&1
\end{array}
\right)
\end{eqnarray}
If we choose ${\mathfrak h}=1+\epsilon$ we find that
\begin{equation}
{\cal E}_0=1;\;\;\;\;\;{\cal E}_1=-1;\;\;\;\;\;{\cal E}_2=-1
\end{equation}
In this case
\begin{eqnarray}
g=
\left (
\begin{array}{cc}
1&-1\\
-1&-1
\end{array}
\right);\;\;\;\;\;
G=
\left (
\begin{array}{cc}
-1&1\\
1&1
\end{array}
\right)
\end{eqnarray}

\section{Finite quantum systems}\label{BBB}
We consider a quantum system with a $d$-dimensional Hilbert space ${\cal H}$. 
A physical example is a system with spin $j$ for which $d=2j+1$.
We consider an orthonormal basis of `position states' which we denote as  
$\ket {{\cal X};m}$ where $m$ takes values in the ring ${\cal Z}_d$.  Here ${\cal X}$ is not a variable, but it simply indicates position states.
Through a Fourier transform we will define the `momentum basis' $\ket {{\cal P};m}$ and the position-momentum phase space.

\subsection{Fourier transforms}

The Fourier transform is defined as:
\begin{eqnarray}\label{Fou}	
{\cal F}=d^{-1/2} \sum_{m=0}^{d-1}  \sum_{n=0}^{d-1}\omega(mn)|{\cal X};m\rangle \langle {\cal X};n|
\end{eqnarray}
where
\begin{eqnarray}
\omega(\alpha)\equiv \omega ^{\alpha}=\exp \left [i \frac{2\pi \alpha}{d}\right ]
\end{eqnarray}
The Fourier transform obeys the relation:
\begin{eqnarray}\label{F}
{\cal F}^4={\bf 1}
\end{eqnarray}
The momentum states are defined as
\begin{equation}\label{PPPT}
|{\cal P};m\rangle={\cal F} |{\cal X};m\rangle=d^{-1/2} \sum_{n} \omega (mn)|{\cal X};n\rangle
\end{equation}
and they form another orthonormal basis in ${\cal H}$.
The position-momentum phase space is the toroidal 
lattice ${\cal Z}_d\times {\cal Z}_d$.

Position and momentum operators ${\hat x}$ and ${\hat p}$ are defined as
\begin{eqnarray}\label{000}
{\hat x}=\sum _{n=0}^{d-1}n|{\cal X};n\rangle \langle {\cal X};n|;\;\;\;\;\;\; 
{\hat p}=\sum _{n=0}^{d-1}n|{\cal P};n\rangle \langle {\cal P};n|;\;\;\;\;\;\;{\hat p}={\cal F} {\hat x} {\cal F}^{\dagger}
\end{eqnarray}

\subsection{Displacements}

In the ${\cal Z}_d\times {\cal Z}_d$ phase space we define the displacement operators
\begin{eqnarray}\label{99}
{\cal Z}&=&\omega ^{\hat x}=\sum _{n=0}^{d-1}\omega (n)|{\cal X};n\rangle \langle {\cal X};n|\nonumber\\
{\cal X}&=&\omega ^{-\hat p}=\sum _{n=0}^{d-1}\omega (-n)|{\cal P};n\rangle \langle {\cal P};n|
\end{eqnarray}
They are displacement operators in the sense that:
\begin{equation}\label{A}
{\cal Z}^\alpha |{\cal P};m \rangle=|{\cal P}; m+\alpha \rangle ;\;\;\;\;\;\;
{\cal Z}^\alpha|{\cal X}; m\rangle= \omega(\alpha m)|{\cal X}; m\rangle
\end{equation}
\begin{equation}\label{B}
{\cal X}^\beta |{\cal P}; m \rangle= \omega(-m\beta)|{\cal P}; m \rangle ;\;\;\;\;\;
{\cal X}^\beta|{\cal X}; m\rangle=|{\cal X}; m+\beta\rangle
\end{equation}
Powers of the displacements operators form the Heisenberg-Weyl group. Indeed we can show that 
the displacement operators obey the relations
\begin{equation}\label{C}
{\cal X}^{d}={\cal Z}^{d}={\bf 1};\;\;\;\;\;\;
{\cal X}^\beta {\cal Z}^\alpha= {\cal Z}^\alpha {\cal X}^\beta \omega(-\alpha \beta)
\end{equation}
where $\alpha$,$\beta$ are integers in ${\cal Z}_d$. 
General displacement operators are defined as
\begin{equation}\label{displac}
{\cal D}(\alpha, \beta)={\cal Z}^\alpha {\cal X}^\beta \omega(-2^{-1}\alpha \beta)
\end{equation}

\section{Quantum systems with dimension $d=p^\ell$}

\subsection{Galois quantum systems}

We have considered quantum systems with position in the ring ${\cal Z}_d$.
When $d=p$ (where $p$ is a prime number) the ${\cal Z}_p$ is a field.
We will use the concept of field extension to introduce quantum systems
with position in the Galois field $GF(p^\ell )$.
In doing so we combine ideas from Galois fields, Fourier transform, quantum mechanics
and applied harmonic analysis.

Let ${\cal H}$ be $p$-dimensional Hilbert space, where $p$ is an odd prime.
We consider the tensor product 
\begin{eqnarray}	
H={\cal H}\otimes...\otimes {\cal H}
\end{eqnarray} 
of $\ell$ such spaces and use
calligraphic letters for operators and states on the various 
$p$-dimensional Hilbert spaces ${\cal H}$;
and ordinary letters for operators and states on the 
$p^\ell$-dimensional Hilbert space $H$.

The position states in $H$ can be labeled with $m\in GF(p^\ell )$ in the following way:
\begin{eqnarray}	
\ket{X;m}&\equiv &\ket{{\cal X};m_0}\otimes ...\otimes \ket{{\cal X};m_{\ell -1}}
\end{eqnarray}
where the notation of Eq.(\ref{kkk}) is used for the components of $m$.  
This is, of course, a trivial labeling rule.
Galois theory ideas enter into quantum systems when  
we introduce the Fourier transform as:
\begin{eqnarray}\label{77777}	
F&=&(p^\ell)^{-1/2} \sum_{m,n} \omega[{\rm Tr}({\mathfrak h}mn)]|X;m\rangle \langle X; n|\nonumber\\
\omega&=& \exp \left (i\frac{2\pi}{p}\right );\;\;\;\;\;
F^4={\bf 1}
\end{eqnarray}
The characters of Eq.(\ref{ppp}) enter in this Fourier transform.
${\mathfrak h}$ plays the role of the inverse Planck's constant.

Acting with this Fourier transform on position states we get momentum states  
\begin{eqnarray}\label{10}
|P; m\rangle&=&F |X; m\rangle=(p^\ell)^{-1/2} \sum_{n} \omega[{\rm Tr}({\mathfrak h}mn)]|X; n\rangle\nonumber\\
&=&\ket{{\cal P};{\bar m}_0}\otimes ...\otimes \ket{{\cal P};{\bar m}_{\ell -1}}
\end{eqnarray}
The dual components ${\bar m}_i$ of $m$ enter in the momentum states, while the components $m_i$ enter in the position states.

We note that 
$|P; m\rangle$ is different from $\ket{{\cal P};m_0}\otimes ...\otimes \ket{{\cal P};m_{\ell -1}}$.
Related to this is the fact that
$F$ is different from the operator ${\cal F}\otimes...\otimes {\cal F}$.
Indeed
\begin{eqnarray}\label{frt}	
F=\sum _m\ket{{\cal P};{\bar m}_0}\bra{{\cal X};m_0}\otimes ...
\otimes \ket{{\cal P};{\bar m}_{\ell -1}}\bra{{\cal X};m_{\ell -1}}
\end{eqnarray}
and
\begin{eqnarray}\label{frt1}	
{\cal F}\otimes...\otimes {\cal F}=\sum _m\ket{{\cal P};m_0}\bra{{\cal X};m_0}\otimes ...
\otimes \ket{{\cal P};m_{\ell -1}}\bra{{\cal X};m_{\ell -1}}
\end{eqnarray}
We stress that a Galois quantum system with Hilbert space $H$,
Fourier transform $F$ and positions (and momenta) in $GF(p^\ell )$
is different from a
quantum system with the same Hilbert space $H$ but with
Fourier transform ${\cal F}\otimes...\otimes {\cal F}$ and positions (and momenta) in
${\cal Z}_p\times...\times {\cal Z}_p$.

The position operator is defined as:
\begin{eqnarray}\label{p1}
{\hat x}=\sum _{m}m|X;m\rangle \langle X;m|=\sum _{\lambda} \epsilon ^\lambda \left [{\bf 1}
\otimes ...\otimes x_{(\lambda)}\otimes ...\otimes {\bf 1}\right ]
\end{eqnarray}
With a Fourier transform we can define the momentum operator as:
\begin{eqnarray}\label{p2}
{\hat p}&=&F{\hat x}F^{\dagger}=\sum _{m}m|P;m\rangle \langle P;m|\nonumber\\&=&\sum _{\lambda}E_{\lambda}\left [{\bf 1}
\otimes ...\otimes p_{(\lambda)}\otimes ...\otimes {\bf 1}\right ]
\end{eqnarray}
Their eigenvalues obey the relation
$m^{p^ \ell}=m$ and therefore according to the Cayley-Hamilton theorem:
\begin{eqnarray}\label{1090}
{\hat x}^{p^\ell}={\hat x};\;\;\;\;\;\;\;{\hat p}^{p^\ell}={\hat p}
\end{eqnarray} 

\subsection{Example}
As an example we consider the $GF(9)$ and we choose the irreducible polynomial $\epsilon ^2+\epsilon +2$,
as in section \ref{AAA1}. We consider the position state
\begin{eqnarray}	
\ket{X;1+\epsilon}=\ket{{\cal X};1}\otimes \ket{{\cal X};1}
\end{eqnarray}
For ${\mathfrak h}=1$, the dual basis is $E_0=-\epsilon$ and  $E_1=-1+\epsilon$. 
Therefore $1+\epsilon=E_0-E_1$ and
\begin{eqnarray}		
\ket{P;1+\epsilon}=F\ket{X;1+\epsilon}=\ket{{\cal P};1}\otimes \ket{{\cal P};-1}
\end{eqnarray}
For ${\mathfrak h}=1+\epsilon$, the dual basis is $E_0=1-\epsilon$ and  $E_1=1+\epsilon$. 
Therefore
\begin{eqnarray}		
\ket{P;1+\epsilon}=F\ket{X;1+\epsilon}=\ket{{\cal P};0}\otimes \ket{{\cal P};1}
\end{eqnarray}

\subsection{Displacements in Galois quantum systems}

We define the displacement operators
\begin{eqnarray}\label{tu0}
Z&=&\sum _{n}\omega [{\rm Tr}({\mathfrak h}n)]|X;n\rangle \langle X;n|\nonumber\\
X&=&\sum _{n}\omega [-{\rm Tr}({\mathfrak h}n)]|P;n\rangle \langle P;n|
\end{eqnarray}
We are interested in more general powers $Z^\alpha$ and $X^\beta$
where $\alpha ,\beta \in GF(p^{\ell})$.
We can easily prove that for $\alpha, \beta \in {\cal Z}_p$ the powers $Z^\alpha$ and $X^\beta$ are given by:
\begin{eqnarray}\label{tu0}
Z^\alpha&=&\sum _{n}\omega [{\rm Tr}({\mathfrak h}\alpha n)]|X;n\rangle \langle X;n|\nonumber\\
X^\beta&=&\sum _{n}\omega [-{\rm Tr}({\mathfrak h}\beta n)]|P;n\rangle \langle P;n|
\end{eqnarray}
But for $\alpha ,\beta \in GF(p^{\ell})$ we need to define the meaning of a complex matrix to a power which belongs in a Galois field.
In this case Eq.(\ref{tu0}) is a definition for $Z^\alpha$ and $X^\beta$.

We next show that
\begin{equation}
Z^\alpha |P;m \rangle=|P; m+\alpha \rangle ;\;\;\;\;\;\;
Z^\alpha|X; m\rangle= \omega[{\rm Tr}({\mathfrak h}\alpha m)]|X; m\rangle
\end{equation}
\begin{equation}
X^\beta |P; m \rangle= \omega[-{\rm Tr}({\mathfrak h}m\beta)]|P; m \rangle ;\;\;\;\;\;
X^\beta|X; m\rangle=|X; m+\beta\rangle
\end{equation}
They are analogous to Eqs(\ref{A}),(\ref{B}) with an extra trace.
We also show 
\begin{equation}\label{33}
X^\beta Z^\alpha= Z^\alpha X^\beta \omega[-{\rm Tr}({\mathfrak h}\alpha \beta)]
\end{equation}
This is analogous to Eq.(\ref{C}) with an extra trace.

General displacement in the $GF(p^{\ell})\times GF(p^{\ell})$ phase space, are defined as:
\begin{eqnarray}\label{90}
D(\alpha, \beta)=Z^\alpha X^\beta \omega\left [-\frac{1}{2}{\rm Tr}({\mathfrak h}\alpha \beta)\right ]
\end{eqnarray}
There is an interesting relation between the displacement operators acting on $H$ and
the displacement operators ${\cal D}$ of Eq. (\ref{displac}) acting on the various $p$-dimensional Hilbert spaces
${\cal H}$:
\begin{eqnarray}\label{p11}
D(\alpha, \beta)&=&{\cal D}({\bar \alpha}_0,\beta _0)\otimes... \otimes{\cal D}({\bar \alpha}_{\ell -1},\beta _{\ell -1})
\end{eqnarray}
Here ${\bar \alpha}_i$ are the dual components of $\alpha$ and $\beta _i$ are the components of $\beta$ as in Eq.(\ref{kkk}).

\section {Frobenius formalism}

\subsection{Frobenius automorphism and irreducible polynomials}

The product 
\begin{eqnarray}\label{irr}
f(y)\equiv (y-\alpha )(y-\alpha ^p)...(y-\alpha ^{p^{\ell-1}})
\end{eqnarray}
involves all the Galois conjugates and is an irreducible polynomial of degree $\ell$ in ${\cal Z}_p[y]$ (the polynomials with coefficients in  ${\cal Z}_p$).

The product of all distinct irreducible polynomials in  ${\cal Z}_p[y]$
of degree $d$, where $d$ is a divisor of $\ell$, is: 
\begin{eqnarray}\label{765}
\prod f_i(y)=y^{p^\ell}-y
\end{eqnarray}
For simplicity, $\ell$ is taken  below to be a prime number.
In this case the ${\cal Z} _p$ is the only proper subfield of $GF(p^\ell)$
and we have a two-layer structure with a total of $p+s$ irreducible polynomials, where
\begin{eqnarray}
s=\frac{p^{\ell}-p}{\ell}.
\end{eqnarray}
The first layer has $s$ irreducible polynomials with degree $\ell$; and to each of them correspond $\ell$ 
Galois conjugates. We label them with $i=0,...,s-1$. 
The second layer has the $p$ polynomials $f_i(y)=y-m$ where $m \in {\cal Z} _p$.
We label them with $i=s,...,s+p-1$. 
Their product is 
\begin{eqnarray}\label{765a}
\prod _{i=0}^{s-1} f_i(y)=\frac{y^{p^\ell}-y}{y^p-y};\;\;\;\;\;\;\prod _{i=s}^{s+p-1} f_i(y) =y^{p}-y.
\end{eqnarray}
The fact that ${\cal Z} _p$ is a subfield of $GF(p^\ell)$ implies that the $y^{p}-y$ divides the $y^{p^\ell}-y$.

\subsection{Frobenius subspaces}

The Frobenius automorphism is central in Galois formalism and in this section we study
its implications in our context. We show that  
the full Hilbert space splits naturally into `Frobenius subspaces'
comprised by states labelled with Galois conjugates.
We also construct transformations which leave these subspaces invariant. 

We call conjugate position states the ones which are labelled with Galois conjugate numbers. 
They are the states
\begin{eqnarray}	
\ket{X;m^{p ^\lambda}}&\equiv &\ket{{\cal X};\sum _{\mu}({\cal C} ^\lambda )_{0, \mu}m _{\mu}}\otimes ...\otimes 
\ket{{\cal X};\sum _{\mu}({\cal C} ^\lambda )_{\ell -1, \mu}m _{\mu}}
\end{eqnarray}
where $\lambda$ takes the values $0,...,\ell -1$.
We split the Hilbert space $H$ into subspaces, each of which is spanned by conjugate position states.
All conjugates correspond to a particular irreducible polynomial and
we label each of these subspaces with the index of the corresponding irreducible polynomial.
For the irreducible polynomial
\begin{eqnarray}\label{8307}
f_i(y)&=&(y-m)(y-m^p)...(y-m^{p^{\ell-1}})
\end{eqnarray}
we consider the space
\begin{eqnarray}\label{830}
H_{Xi}&=&{\rm span}\{\ket{X;m},\ket{X;m ^p},...,\ket{X;m ^{p^{\ell -1}}}\}
\end{eqnarray}
The index $i$ indicates the corresponding irreducible polynomial and the
index $X$ indicates that the position states have been used.
We can use the states $U\ket{X;m}$ where $U$ is any unitary operator, and we will get 
different Frobenius subspaces which we denote as $H_{UXi}$.
We call $\Pi_{Xi}$ the projection operator to the space $H_{Xi}$.

The Hilbert space $H$ is the direct sum of all Frobenius subspaces.
For prime $\ell $, there are $s$  Frobenius subspaces which are $\ell $-dimensional and we call $H_A$ their direct sum; and
there are $p$ Frobenius subspaces which are one-dimensional and we call $H_B$ their direct sum:
\begin{eqnarray}\label{p09}
H=H_A\bigoplus H_B;\;\;\;\;\;\;H_A=\bigoplus _{i=0}^{s-1} H_{Xi};\;\;\;\;\;\;H_B=\bigoplus _{i=s}^{s+p-1} H_{Xi}
\end{eqnarray}

\subsection{Frobenius transformations}

We call Frobenius transformations the following unitary transformations in $H_{Xi}$:
\begin{eqnarray}\label{45}
{\cal G}_i\equiv \sum _{\lambda \in {\cal Z}_{\ell} }|X;m^{p^{\lambda +1}}\rangle \langle X; m^{p^{\lambda }}|
\end{eqnarray}
Here $m$ is one of the Galois conjugates corresponding to the irreducible polynomial $f_i(y)$
and by taking all $\lambda \in {\cal Z}_{\ell}$ we get all the Galois conjugates.
For the one dimensional spaces $H_{Xi}$, the ${\cal G}_i$ is simply the projection operator $\Pi_{Xi}$:
\begin{eqnarray}\label{94v}
{\cal G}_i=\Pi_{Xi};\;\;\;\;\;i=s,...,s+p-1
\end{eqnarray}
We sum all the transforms ${\cal G}_i$ and we get 
\begin{eqnarray}
{\cal G}=\sum _{i=0}^{s+p-1} {\cal G}_i
\end{eqnarray}
We can show that
\begin{eqnarray}
{\cal G}^\ell={\bf 1};\;\;\;\;\;
[{\cal G},\Pi_{Xi}]=0
\end{eqnarray}
The operators  ${\cal G} ^\lambda $ map the position state $|X;m\rangle$ to its Galois conjugate position states $|X;m^{p^\lambda}\rangle$.
We can express this in terms of the various components of these states as:
\begin{eqnarray}
&&{\cal G}^ \lambda \ket{{\cal X};m_0}\otimes ...\otimes \ket{{\cal X};m_{\ell -1}}\nonumber \\&&=
\ket{{\cal X};\sum _{\mu}({\cal C} ^\lambda )_{0 ,\mu}m _{\mu}}\otimes ...
\otimes \ket{{\cal X};\sum _{\mu}({\cal C} ^\lambda )_{\ell -1, \mu}m _{\mu}}
\end{eqnarray}

We can show that
\begin{eqnarray}\label{890}
{\cal G}^\lambda X^{\beta}({\cal G}^{\dagger})^\lambda=X^{\beta ^{p^\lambda}};\;\;\;\;\;\;\;
{\cal G}^\lambda Z^{\alpha}({\cal G}^{\dagger})^\lambda=Z^{\alpha ^{p^\lambda}{\mathfrak h}^{p^\lambda-1}}
\end{eqnarray}
and more generally that
\begin{eqnarray}\label{890d}
{\cal G}^\lambda D(\alpha, \beta)({\cal G}^{\dagger})^\lambda&=&D\left (\alpha ^{p^\lambda}{\mathfrak h}^{p^\lambda-1}, \beta ^{p^\lambda}\right )
\end{eqnarray}
where $\lambda \in {\cal Z}_{\ell}$.
When $\alpha, \beta \in {\cal Z}_p$ the ${\cal G}$ commutes with  $D(\alpha, \beta)$.
We note here that although ${\cal G}$ commutes with $X$ and $Z$, it does {\bf not} commute with its powers
$X^\alpha$ and $Z^\alpha$ when $\alpha \in GF(p^\ell)$. 

We next use Eq.(\ref{777}) and the relation
\begin{eqnarray}
\alpha ^{p^\lambda}{\mathfrak h}^{p^\lambda-1}=\sum _{\mu}A_{\mu}E _{\mu}
;\;\;\;\;\;\;A_{\mu}\equiv \sum _{\kappa }{\bar \alpha }_{\kappa}({\cal C} ^{- \lambda} )_{\kappa \mu}
\end{eqnarray}
and taking into account Eq.(\ref{p11}) we rewrite Eq.(\ref{890d}) as:
\begin{eqnarray}\label{p110}
&&{\cal G}^\lambda{\cal D}({\bar \alpha}_0,\beta _0)\otimes... \otimes{\cal D}({\bar \alpha}_{\ell -1},\beta _{\ell -1})({\cal G}^{\dagger})^\lambda
=\nonumber\\
&&{\cal D}\left (A_0,\sum _{\mu}({\cal C} ^\lambda )_{0,\mu}\beta _{\mu}\right )
\otimes... \otimes{\cal D}\left (A_{\ell -1},\sum _{\mu}({\cal C} ^\lambda )_{\ell -1,\mu}\beta _{\mu}\right )
\end{eqnarray}

\subsection{Quantum formalism in ${\mathfrak H}_{X\kappa}$}

In this section we use the notation $m(i)$ for one of the conjugates corresponding to the irreducible polynomial $f_i(y)$.
The rest of the conjugates corresponding to the same polynomial are $[m(i)]^p,...,[m(i)]^{p^{\ell -1}}$.

The Hilbert space $H_A$ in Eq.(\ref{p09}) can be written as the direct sum of the spaces ${\mathfrak H}_{X\kappa}$
which are defined as follows. We take
one position state from each of the $\ell$-dimensional spaces $H_{Xi}$ in Eq.(\ref{830})
and introduce the $s$-dimensional space:
\begin{eqnarray}
{\mathfrak H}_{X0}&=&{\rm span}\{\ket{X;m(0)},...,\ket{X;m(s-1)}\}
\end{eqnarray}
Since each of the $m(0),...,m(s-1)$ corresponds to a different irreducible polynomial, there are no Galois conjugates among them.
Acting with powers of ${\cal G}$ on ${\mathfrak H}_{X0}$ we get the following `copies' of ${\mathfrak H}_{X0}$:  
\begin{eqnarray}
{\mathfrak H}_{X\kappa}&=&{\rm span}\{\ket{X;[m(0)]^{p ^\kappa}},...,\ket{X;[m(s-1)]^{p ^\kappa}}\}
\end{eqnarray}
It is easily seen that
\begin{eqnarray}
H_A=\bigoplus _{\kappa =0}^{\ell -1} {\mathfrak H}_{X\kappa}
\end{eqnarray} 
We call $\Sigma _{X\kappa}$ the projection operator to the space ${\mathfrak H}_{X\kappa}$.

We construct briefly a quantum formalism analogous to the one presented in section \ref{BBB} for each of the  $s$-dimensional spaces ${\mathfrak H}_{X\kappa}$.
Here the states $|X;[m (\lambda)]^{p^\kappa}\rangle$ play the role of `position states' labeled with $\lambda \in {\cal Z}_s$; 
the dual (`momentum') states are defined through a Fourier transform $\mathfrak{F}_\kappa$ which is defined below; 
and the phase space is ${\cal Z}_s \times {\cal Z}_s$.

In ${\mathfrak H}_{X\kappa}$ we introduce the Fourier transform:
\begin{eqnarray}\label{94v}
&&\mathfrak{F}_{\kappa}=s ^{-1/2}\sum _{\lambda ,\mu}\Omega (\lambda \mu)|X;[m(\lambda)]^{p^\kappa}\rangle \langle X; [m(\mu)]^{p^\kappa}|
;\;\;\;\;\lambda ,\mu \in {\cal Z}_s\nonumber\\
&&\Omega(\lambda )\equiv \Omega ^{\lambda }=\exp \left [i \frac{2\pi \lambda}{s}\right ];\;\;\;\;\;\;\;\mathfrak{F}_\kappa ^4={\Sigma} _{X\kappa}
\end{eqnarray}
Clearly the Fourier transform $\mathfrak{F}_\kappa$ is very different from the Fourier transform $F$ of Eq.(\ref{77777}).

We define displacement operators in ${\mathfrak H}_{X\kappa}$ as
\begin{eqnarray}\label{4500}
&&{\cal S}_\kappa =\sum _{\lambda \in {\cal Z}_s}|X;[m(\lambda +1)]^{p^\kappa}\rangle \langle X; [m(\lambda)]^{p^\kappa}|\nonumber\\
&&{\cal R}_\kappa =\mathfrak{F}_\kappa {\cal S}_\kappa \mathfrak{F}_\kappa ^{\dagger}=
\sum _{\lambda \in {\cal Z}_s}\Omega ^{\lambda}|X;[m(\lambda )]^{p^\kappa}\rangle \langle X; [m(\lambda)]^{p^\kappa}|
\end{eqnarray}
In analogy with Eqs(\ref{C}), we can show that the ${\cal S}_\kappa$, ${\cal R}_\kappa $ form a Heisenberg-Weyl group:
\begin{eqnarray}
&&{\cal S}_\kappa ^{\lambda}{\cal R}_\kappa ^{\mu}={\cal R}_\kappa ^{\mu}{\cal S}_\kappa ^{\lambda}\Omega (-\lambda \mu);\;\;\;\;\;\;
\lambda, \mu \in {\cal Z}_s\nonumber\\
&&{\cal S}_\kappa ^s={\cal R}_\kappa ^s=\Sigma _{X\kappa}
\end{eqnarray}

We act with $\mathfrak{F}_\kappa $ on the $\ell$ position states of Eq.(\ref{830}) and we get the dual states:
\begin{eqnarray}\label{94vi}
\ket{\mathfrak P;[m(\lambda) ^{p^\kappa }}=\mathfrak{F}_\kappa \ket{X;[m(\lambda )] ^{p^\kappa }}=
s ^{-1/2}\sum _{\mu \in {\cal Z}_s}\Omega (\lambda \mu)|X;[m(\lambda )]^{p^\kappa}\rangle
\end{eqnarray}
They can be viewed as `momentum states' within the space ${\mathfrak H}_{X\kappa}$; but they are very different from the 
states $\ket{P;m}=F\ket{X;m}$.
In analogy to Eqs(\ref{A}),(\ref{B}) we get:
\begin{eqnarray}\label{A1}
&&{\cal S}_\kappa \ket{X;[m (\lambda)]^{p^\kappa }}=\ket{X;[m(\lambda +1)] ^{p^\kappa }}\nonumber\\
&&{\cal S}_\kappa \ket{\mathfrak P;[m(\lambda)] ^{p^\kappa }}=\Omega ^{-\lambda}\ket{\mathfrak P;[m(\lambda)] ^{p^\kappa }} 
\end{eqnarray}
and also
\begin{eqnarray}\label{B1}
&&{\cal R}_\kappa \ket{X;[m (\lambda)]^{p^\kappa }}= \Omega ^{\lambda}\ket{X;[m(\lambda)] ^{p^\kappa }}\nonumber\\
&&{\cal R}_\kappa \ket{\mathfrak P;[m (\lambda)]^{p^\kappa }}=\ket{\mathfrak P;[m (\lambda +1)]^{p^{\kappa}}}
\end{eqnarray}
We also introduce the `position' operator in $\mathfrak{H}_{X\kappa}$ 
\begin{eqnarray}
{\mathfrak r}_\kappa =\sum _{\lambda \in {\cal Z}_s}\lambda \ket{X;[m(\lambda )] ^{p^\kappa }}\bra{X;[m(\lambda)] ^{p^\kappa }}
\end{eqnarray}
and its dual
\begin{eqnarray}
{\mathfrak g}_\kappa &=&\mathfrak{F}_\kappa {\mathfrak g}_\kappa \mathfrak{F}_\kappa ^{\dagger}
=\sum _{\lambda \in {\cal Z}_s}\lambda \ket{\mathfrak P;[m (\lambda)]^{p^\kappa }}\bra{\mathfrak P;[m (\lambda)]^{p^\kappa }}\nonumber\\&=&
\frac{1}{2\pi is}\sum _{\lambda, \mu}\Delta _1 (\lambda -\mu)\ket{X;[m(\lambda)] ^{p^\kappa }}\bra{X ;[m(\mu)] ^{p^\kappa }}
\end{eqnarray}
The function $\Delta _1 (\lambda )$ is defined as
\begin{eqnarray}
\Delta _1 (\lambda )=2\pi i\sum _{m=0}^{s-1}m\Omega (m\lambda)
\end{eqnarray}
It is explained in \cite{rev} that it is the analogue of the first derivative
of delta function in the present context.
The operators ${\mathfrak r}_\kappa $ and ${\mathfrak g}_\kappa $ are very different from the position and momentum operators $\hat x$
and $\hat p$ in Eqs(\ref{p1}),(\ref{p2}).
In analogy with Eq.(\ref{99}) we show that
\begin{eqnarray}
{\cal R}_\kappa =\Omega ^{{\mathfrak r}_\kappa };\;\;\;\;\;\;{\cal S}_\kappa =\Omega ^{-{\mathfrak g}_\kappa }
\end{eqnarray} 

We note that acting with ${\cal G}$ on both sides of the various operators that we defined above in ${\mathfrak H}_{X\kappa}$
we get the corresponding operators in ${\mathfrak H}_{X\kappa +1}$:
\begin{eqnarray}
&&{\cal G}{\mathfrak r}_\kappa {\cal G}^{\dagger}={\mathfrak r}_{\kappa +1};\;\;\;\;\;\;{\cal G}{\mathfrak g}_\kappa {\cal G}^{\dagger}
={\mathfrak g}_{\kappa +1}\nonumber\\
&&{\cal G}{\cal R}_\kappa {\cal G}^{\dagger}={\cal R}_{\kappa +1};\;\;\;\;\;\;{\cal G}{\cal S}_\kappa {\cal G}^{\dagger}={\cal S}_{\kappa +1}
\end{eqnarray}

Physically, the transformations ${\cal S}_\kappa$ can be implemented in a system with Hamiltonian ${\mathfrak g}_\kappa$.
In this case the evolution operator $\exp (-it {\mathfrak g}_\kappa)$ becomes equal to a power of ${\cal S}_\kappa$ at times
which are integral multiples of $t_0=2\pi /s$. In such a system,  
the above formalism provides information about the `stroboscopic evolution'.
For example a state which belongs in the subspace ${\mathfrak H}_\kappa $ will evolve at times $Nt_0$ (where $N$ is an integer) 
into another state in the same subspace.
Also, the system will return to its original states at times which are integer multiples of $Ns t_0$.
Similar comments can be made for the transformations ${\cal R}_\kappa$ which can be implemented in a system with Hamiltonian ${\mathfrak r}_\kappa$.

The use of the spaces ${\mathfrak H}_\kappa$in real physical problems requires further work.
Potential applications include quantum coding, other schemes to protect quantum 
devices from enviromental noise, quantum information processing, the magnetic translation group in condensed matter, quantum chaos, etc.

\subsection{Example}

As an example we consider the $GF(9)$ and choose the irreducible polynomial $\epsilon ^2+\epsilon +2$. 
The various irreducible polynomials are in this case factorized as follows:
\begin{eqnarray}\label{6ty}
&&f_0(y)=y^2+2y+2=(y-1-\epsilon)(y-2\epsilon)\nonumber\\
&&f_1(y)=y^2+y+2=(y-\epsilon)(y-2-2\epsilon)\nonumber\\
&&f_2(y)=y^2+1=(y-1-2\epsilon)(y-2-\epsilon)\nonumber\\
&&f_3(y)=y\nonumber\\
&&f_4(y)=y-1\nonumber\\
&&f_5(y)=y-2\nonumber\\
\end{eqnarray}
Eqs.(\ref{765a}),become in this case
\begin{eqnarray}
&&f_0(y)f_1(y)f_2(y)=\frac{y^9-y}{y^3-y}\nonumber\\
&&f_3(y)f_4(y)f_5(y)=y^3-y
\end{eqnarray}

The Hilbert space $H$ splits into six Frobenius subspaces 
\begin{eqnarray}\label{p091}
H=H_A\bigoplus H_B;\;\;\;\;\;\;
H_A=\bigoplus _{i=0}^{2} H_{Xi};\;\;\;\;\;\;H_B=\bigoplus _{i=3}^{5} H_{Xi}
\end{eqnarray} 
where
\begin{eqnarray}\label{po9}
&&H_{X0}={\rm span}\{\ket {X;1+\epsilon},\ket {X;2\epsilon}\}\nonumber\\
&&H_{X1}={\rm span}\{\ket {X;\epsilon},\ket {X;2+2\epsilon}\}\nonumber\\
&&H_{X2}={\rm span}\{\ket {X;1+2\epsilon},\ket {X;2+\epsilon}\}\nonumber\\
&&H_{X3}={\rm span}\{\ket {X;0}\}\nonumber\\
&&H_{X4}={\rm span}\{\ket {X;1}\}\nonumber\\
&&H_{X5}={\rm span}\{\ket {X;2}\}
\end{eqnarray}
The operator ${\cal G}$ is given by:
\begin{eqnarray}
{\cal G}=\sum _{i=0}^5 {\cal G}_i
\end{eqnarray}
where
\begin{eqnarray}
{\cal G}_0&=&\ket {X;1+\epsilon}\bra {X;2\epsilon}+\ket {X;2\epsilon}\bra {X;1+\epsilon}\nonumber\\
{\cal G}_1&=&\ket {X;\epsilon}\bra {X;2+2\epsilon}+\ket {X;2+2\epsilon}\bra {X;\epsilon}\nonumber\\
{\cal G}_2&=&\ket {X;1+2\epsilon}\bra {X;2+\epsilon}+\ket {X;2+\epsilon}\bra {X;1+2\epsilon}\nonumber\\
{\cal G}_3&=&\ket {X;0}\bra {X;0}\nonumber\\
{\cal G}_4&=&\ket {X;1}\bra {X;1}\nonumber\\
{\cal G}_5&=&\ket {X;2}\bra {X;2}
\end{eqnarray}

The spaces ${\mathfrak H}_{X0}$ and ${\mathfrak H}_{X1}$ are
\begin{eqnarray}
{\mathfrak H}_{X0}&=&{\rm span}\{ \ket {X;1+\epsilon},\ket {X;\epsilon},\ket {X;1+2\epsilon}  \}\nonumber\\
{\mathfrak H}_{X1}&=&{\rm span}\{\ket {X;2\epsilon},\ket {X;2+2\epsilon},\ket {X;2+\epsilon}\}\nonumber\\
H_A&=&{\mathfrak H}_{X0}\bigoplus {\mathfrak H}_{X1}
\end{eqnarray}
In the space ${\mathfrak H}_{X0}$ the operators ${\cal S}_0$ and ${\cal R}_0$ are given by
\begin{eqnarray}
{\cal S}_0&=&\ket {X;\epsilon}\bra{X;1+\epsilon}+\ket {X;1+2\epsilon}\bra {X;\epsilon}\nonumber\\&+&\ket {X;1+\epsilon}\bra {X;1+2\epsilon}\nonumber\\
{\cal R}_0&=&\ket {X;1+\epsilon}\bra{X;1+\epsilon}+\Omega \ket {X;\epsilon}\bra {X;\epsilon}\nonumber\\
&+&\Omega ^2 \ket {X;1+2\epsilon}\bra {X;1+2\epsilon}\nonumber\\
\Omega&=&\exp \left (i\frac{2\pi}{3}\right )
\end{eqnarray}
In the space ${\mathfrak H}_{X1}$ the operators ${\cal S}_1$ and ${\cal R}_1$ are given by
\begin{eqnarray}
{\cal S}_1&=&\ket {X;2+2\epsilon}\bra{X;2\epsilon}+\ket {X;2+\epsilon}\bra {X;2+2\epsilon}\nonumber\\&+&\ket {X;2\epsilon}\bra {X;2+\epsilon}\nonumber\\
{\cal R}_1&=&\ket {X;2\epsilon}\bra{X;2\epsilon}+\Omega \ket {X;2+2\epsilon}\bra {X;2+2\epsilon}\nonumber\\
&+&\Omega ^2 \ket {X;2+\epsilon}\bra {X;2+\epsilon}
\end{eqnarray}

\section{Frobenius formalism and Riemann surfaces}

The Frobenius formalism led to $\ell$ copies of the space ${\mathfrak H}_{X0}$.
This is similar concept to the $\ell$ sheets in a Riemann surface related to the map 
\begin{eqnarray}
z\;\rightarrow\;z^{1/\ell }
\end{eqnarray}
More generally, there are links between the Frobenius formalism in Galois theory and Riemann surfaces.
They have been discussed in various contexts (e.g., \cite{R}) and here we discuss these links in our own context.

Various analytic representations have been used in quantum mechanics (for a review see \cite{V06}).
Here we show that an analytic representation of quantum states in the space $H_A$ can be defined 
in the $\ell$-sheeted covering of a sphere, with states in ${\mathfrak H}_{X\kappa} $ represented by 
functions on the $\kappa$-sheet. This analytic representation is
similar to the one discussed in \cite{VVV} in a model which 
was introduced from a Riemann surfaces point of view and did not involve Galois theory.
The aim here is to show the conceptual link of this model to Galois theory.
The model uses $SU(2)$ transformations and $SU(2)$ coherent states in the spaces ${\mathfrak H}_{X\kappa }$.
We refer to \cite{VVV} for technical details and here we present briefly only a few formulas which show the
connection between Riemann surfaces and  Galois theory in a quantum mechanical context.
For simplicity we consider the Bose case of odd $s$; but the formalism can be extended to the Fermi case of even $s$ also.

We consider a sphere where a point is described in spherical coordinates with the angles $(\alpha, \beta)$,
where $0\le \alpha\le \pi$, $0\le \beta < 2\pi$.
A sphere is topologically equivalent to the extended complex plane $C_E=C\cup \{\infty \}$ and 
the stereographic projection 
\begin{eqnarray}\label{stereo}
z=-\tan \left ( \frac {\alpha}{2} \right )e^{-i\beta}.
\end{eqnarray}
provides a one-to-one mapping between the two.
The south pole is mapped to the point $z=0$ and the north pole to $\infty$. The metric in the extended complex plane is 
\begin{eqnarray}\label{lkj}
d\mu (z)=\frac{dz_Rdz_I}{(1+|z|^2)^2}
\end{eqnarray}
where $z_R,z_I$ are the real and imaginary parts of $z$, correspondingly.

We next introduce the $\ell$-sheeted extended complex plane 
as follows. The north and south poles are branch points of order $\ell -1$.
The cuts ${\mathfrak C}_\kappa $ and the sheets ${\mathfrak S}_\kappa$ are given by
\begin{eqnarray}
{\mathfrak C}_\kappa&=&\left \{z=r\vartheta ^\kappa;\;\;r\ge 0\right \};\;\;\;
\vartheta=\exp \left (i\frac{2\pi}{\ell}\right )\nonumber\\
{\mathfrak S}_\kappa&=&\left \{z=r\exp (i\phi);\;\;r\ge 0;\;\;\;\;\;\frac{2\pi \kappa}{\ell}<\phi <\frac{2\pi (\kappa +1)}{\ell}\right \}
\end{eqnarray}
The sheet number of a complex number $z$ is defined as
\begin{eqnarray}
\tau (z)={\rm IP}\left (\frac{\ell \arg (z)}{2\pi}\right);\;\;\;\;\;\tau (z)\in {\cal Z}_\ell
\end{eqnarray}
where IP stands for the integer part of the number. 
In order to find the metric in the $\ell$-sheeted extended complex plane, we replace $z$ with $z^\ell$ in Eq.(\ref{lkj})and we get
\begin{eqnarray}
d\mu _{\ell}(z)=\frac{\ell ^2|z|^{2(\ell -1)}}{(1+|z|^{2\ell})^2}dz_Rdz_I
\end{eqnarray}

A state
\begin{eqnarray}
\ket {f}=\sum _{\lambda,\kappa}f(\lambda,\kappa)|X;[m(\lambda)]^{p^\kappa}\rangle;\;\;\;\;\;\;\sum _{\lambda,\kappa}|f(\lambda,\kappa)|^2=1
\end{eqnarray}
in $H_A$, is represented with the polynomial
\begin{eqnarray}
f(z)=\sum _{\lambda =0}^{s-1} d(s,\lambda)f(\lambda, \tau(z))z^{\ell \lambda}
\end{eqnarray}
where
\begin{eqnarray}
d(s,\lambda)=\left [\frac{(2j)!}{(j+n)!(j-n)!}\right ]^{1/2};\;\;\;\;\;\;j=\frac{s-1}{2}
\end{eqnarray}
The function $f(z)$ in the sheet ${\mathfrak S}_\kappa$ depends only on the projection 
$\Sigma _{X\kappa}\ket {f}$ of the state $\ket {f}$ in the space ${\mathfrak H}_{X\kappa}$. 
A state which belongs entirely in the space ${\mathfrak H}_{X\lambda}$ is represented by a function $f(z)$
which is equal to zero in all sheets apart from the ${\mathfrak S}_\lambda$.

The function 
$f(z)$ is analytic in the interior of all sheets ${\mathfrak S}_\kappa$
and has discontinuities across the cuts ${\mathfrak C}_\kappa$ given by
\begin{eqnarray}
\Delta _\kappa (z)=\sum _{\lambda =0}^{s-1} d(s,\lambda)[f(\lambda, \kappa)-f(\lambda, \kappa -1)]z^{\ell \lambda}
\end{eqnarray}
The scalar product is given by
\begin{eqnarray}
\langle g\ket {f}=\frac{s}{\pi}\int _{C_E} [g(z)]^*f(z)(1+|z|^{2\ell})^{1-s}d\mu _{\ell}(z)
\end{eqnarray}
The Frobenius transformations ${\cal G}$ are easily implemented in this representation as follows. If $f(z)$ represents the state $\ket {f}$ then
\begin{eqnarray}
{\cal G}^\kappa \ket {f}\;\rightarrow\; f(z\vartheta ^\kappa)
\end{eqnarray}

The above analytic representation shows that there is an interesting relation between the Frobenius formalism in Galois theory and Riemann surfaces.
More work is required in this direction, with more complicated quantum systems and more complex Riemann surfaces.

\section{Discussion}

Galois fields have been introduced in quantum mechanics
with a clear physical motivation:
in order to have well defined symplectic transformations;
or in the context of mutually unbiased bases; or in the context of quantum coding.

The motivation in this article is more mathematical and aims to bring a concept similar to field extension in the context of Hilbert spaces.
Field extension constructs large fields from smaller ones.
We started with a `small' Hilbert space ${\cal H}$ describing a system where the position and momentum take values in the field ${\cal Z}_p$.
Through tensor product we constructed
a `large' Hilbert space $H$ with the Fourier transform $F$ of Eq.(\ref{frt}), 
describing a system where the position and momentum take values in the Galois field $GF(p^\ell )$.
This Galois quantum system is different from another
quantum system with the same Hilbert space $H$ but with
Fourier transform ${\cal F}\otimes...\otimes {\cal F}$ of Eq.(\ref{frt1}) and positions (and momenta) in
${\cal Z}_p\times...\times {\cal Z}_p$.

Galois quantum system `inherit' properties from Galois fields and in this paper we discussed the Frobenius formalism.
We have considered the case of prime $\ell$ with a simple two-layer structure; but the discussion can be extended to all $\ell$
with a more complex multi-layer structure. 
We have introduced Frobenius subspaces spanned by Galois conjugate position states and we have
constructed the Frobenius transformations of Eq.(\ref{45}) which leave invariant these subspaces.

We have also introduced the spaces ${\mathfrak H}_{X,0},...,{\mathfrak H}_{X,\ell-1}$ (which are copies of each other) and
constructed explicitly Heisenberg-Weyl groups of transformations in them.
An analytic representation has been used to show the relationship of this construction to $\ell$-sheeted Riemann surfaces. 
Potential applications include quantum coding, quantum information processing, the magnetic translation group in condensed matter, quantum chaos, etc.

In Galois quantum systems we have blended the areas of Galois fields 
and quantum mechanics. 
This is interesting from a mathematical point of view; 
and at the same time it might have several applications.

\end{article}


\begin{thebibliography}{99}





\bibitem{1}
	H. Weyl, Theory of Groups and Quantum Mechanics (Dover, New York, 1950)
\bibitem{2}	
	J. Schwinger, Proc. Nat. Acad. Sci. U.S.A. 46, 570 (1960); Quantum Kinematics and Dynamics (Benjamin, 
	New York, 1970).

\bibitem{3}	L. Auslander, R. Tolimieri Bull. Am. Math.Soc. 1, 847 (1979)

\bibitem{4}	J.H. Hannay, M.V. Berry, Physica 1D, 267 (1980)

\bibitem{5}	 R. Balian and C. Itzykson, C.R. Acad. Sci. 303, 773 (1986)

\bibitem{6}   D.B. Fairlie, P. Fletcher, C.K. Zachos, J. Math. Phys. 31, 1088 (1990)

\bibitem{7}	A. Vourdas, Phys. Rev. A41, 1653 (1990)\\
 	A. Vourdas, Phys. Rev. A43, 1564 (1991)\\
	A. Vourdas, C. Bendjaballah, Phys.Rev. A47, 3523 (1993)

\bibitem{8}
A. Vourdas, J.Phys.A29, 4275 (1996)\\
A. Vourdas, J.Phys.A38, 8453 (2005)

\bibitem{9}	M. Neuhauser, J. Lie Theory 12, 15 (2002)	

\bibitem{10} 	J.P. Paz, Phys. Rev. A65, 062311 (2002)

\bibitem{rev}	A. Vourdas, Rep. Prog. Phys. 67, 267 (2004)

\bibitem{fg}	J.W.P. Hirschfeld, `Projective geometries over finite fields' 
	(Oxford University Press, Oxford, 1979)\\
	L.M. Batten, `Combinatorics of finite geometries' (cambridge University Press, Cambridge, 1997)
\bibitem{m0}
I.D. Ivanovic, J. Phys. A14, 3241 (1981)

\bibitem{m1}
W. Wootters, Ann. Phys. (NY), 176, 1 (1987)\\
W. Wootters, B.D. Fields, Ann. Phys. (NY), 191, 363 (1989)\\
K. Gibbons, M.J. Hoffman, W. Wootters, Phys. Rev. A70, 062101 (2004)

\bibitem{m2}   S. Bandyopadhyay, P.O. Boykin, V.Roychowdhury, F. Vatan, Algorithmica 34, 512 (2002)

\bibitem{m3}	A.O. Pittenger, M.H. Rubin, Linear Algebra Appl. 390, 255 (2004)\\
	A.O. Pittenger, M.H. Rubin, J. Phys. A38, 6005 (2005)

\bibitem{m4}	A. Klappenecker, M. Rotteler, Lect. Notes Comp. Science 2948, 137 (2004)

\bibitem{m5}	P. Wocjan, T. Beth, Quantum Inf. Comput. 5, 181 (2005)

\bibitem{m6}   A. Klimov, L. Sanchez-Soto, H. de Guise, J. Phys. A38, 2747 (2005)\\
A. Klimov, L. Sanchez-Soto, H. de Guise, J. Opt. B:Quantum Semiclass. Opt. 7, 283 (2005)\\
J.L. Romero, G. Bjork, A.B. Klimov, L.L. Sanchez-Soto, Phys. Rev. A72, 062310 (2005)

\bibitem{m7}	 M. Saniga, M. Planat, H. Rosu, J. Opt. B-Quantum Semiclass. Optics 6, L19 (2004)\\	
M. Saniga,	M. Planat, J. Phys. A39, 435 (2006)

\bibitem{m8}	T. Durt, J. Phys. A38, 5267 (2005)\\
S. Colin, J. Corbett, T. Durt, D. Gross,  J. Opt. B-Quantum Semiclass. Optics 7, S778 (2005)

\bibitem{m9}	E.F. Calvao, Phys. Rev. A71, 042302 (2005)

\bibitem{m10}
I. Bengtsson, A. Ericsson, Open Syst. Inf. Dyn. 12, 107 (2005)



\bibitem{mkp1}	B.G. Englert, Y. Aharonov, Phys. Lett. A284, 1 (2001)

\bibitem{mkp2}
A. Hayashi, M. Horibe, T. Hashimoto, Phys. Rev. A71, 052331 (2005)


\bibitem{c1}	A. Asikhmin, E. Knill, IEEE Trans. Inf. Theo., 47, 3065 (2001)
 
\bibitem{c2}	H. Barnum, C. Crepeau, D. Gottesman, A. Smith, A. Tapp, Proceedings of the 43th Annual Symposium on Foundations of Computer Science
(FOCS) (IEEE Computer Society, Los Alamitos, CA, 2002) pp 449-458

\bibitem{c3}	A. Vourdas, Phys. Rev. A65, 042321 (2002)\\
	A. Vourdas, J. Phys. A37, 3305 (2004)	

\bibitem{R}
E. Reyssat in `From Number Theory to Physics'
Ed. M. Waldschmidt, P. Moussa, J.M. Louck, C. Itzykson,
(Springer, Berlin, 1992)

\bibitem{V06}
A. Vourdas, J. Phys. A. 39, R65 (2006)

\bibitem{VVV}
A. Vourdas, J. Math. Phys. 36, 4757 (1995)\\

\end{thebibliography}
\end{document}